\documentclass[12pt,twoside]{article}
\usepackage{graphicx}
\usepackage{cmp2em}
\usepackage{cite}	% sorts references

\date{July 20, 2005}
\title{Entropy of fermionic models on highly frustrated lattices}
\author{A.~Honecker\refaddr{BS}, J.~Richter\refaddr{MD}}
\addresses{
\addr{BS} Technische Universit\"at Braunschweig,
Institut f\"ur Theoretische Physik,
Mendelssohnstrasse 3, 38106 Braunschweig, Germany
\addr{MD} Institut f\"ur Theoretische Physik, Otto-von-Guericke
 Universit\"at Magdeburg,
 39016 Magdeburg, Germany}
\begin{document}

\maketitle

\begin{abstract}
Spinless fermions on highly frustrated lattices are characterized
by a lowest single-particle band which is completely flat.
Concrete realizations are provided by the sawtooth chain
and the kagom\'e lattice. For these models a real-space picture
is given in terms of localized states. Furthermore, we find
a finite zero-temperature entropy for a suitable choice of the
chemical potential. The entropy is computed numerically at
finite temperature and one observes a strong cooling effect
during adiabatic changes of the chemical potential. We argue
that the localized states, the associated zero-temperature
entropy and thus also the large temperature variations carry
over to the repulsive Hubbard model.
The relation to flat-band ferromagnetism is also discussed briefly.

% Up to six keywords
%
\pacs
% Up to six PACS numbers
71.10.Fd; % Lattice fermion models (Hubbard model, etc.)
65.40.Gr; % Entropy and other thermodynamical quantities
75.30.Sg; % Magnetocaloric effect, magnetic cooling
75.10.Jm  % Quantized spin models

\end{abstract}

\section{Introduction}

Frustrated quantum magnets exhibit a rich variety of semi-classically
ordered and disordered ground states. This has been analyzed
in some detail for two-dimensional models (see \cite{lhuillier03,RSH04}
for recent reviews).
Interesting behaviour is also observed in a magnetic field,
including plateaux in the zero-temperature magnetization curve
and field-induced quantum phase transitions
(compare \cite{jump,RSH04,RSHSS04} and references therein).
A particularly intriguing phenomenon are non-interacting
localized magnons which have recently been discovered
at the saturation field of the spin-$s$ $XXZ$ model
on highly frustrated lattices \cite{jump,RSH04,RSHSS04,SSRS01,RiLoc}.
This gives rise to an enhanced magnetocaloric effect. Indeed,
cooling rates around the saturation field in geometrically frustrated
classical spin systems were predicted \cite{Zhito} to be up to
several orders of magnitude bigger than in non-frustrated spin models.
Enhanced cooling rates have also been verified experimentally,
namely for Gd$_3$Ga$_5$O$_{12}$, a hyper-kagom\'e lattice (see
\cite{ZGBP} and references therein), and the pyrochlore magnet
Gd$_2$Ti$_2$O$_7$ \cite{SPSGBPBZ}. Since these results suggest applications
for efficient low-temperature magnetic refrigeration, the magnetocaloric
properties of the corresponding quantum systems are currently
under intense investigation \cite{ZhiHo,DerRi,ZhiTsu,Schnack,ZhiTsuP,HoWe}.

With the aid of the Jordan-Wigner transformation (see e.g.\ Chapter 1 of
\cite{takahashi}), $s=1/2$ $XXZ$ chains can be mapped
to spinless fermions. These are generally interacting, but the interaction
is irrelevant
at low densities for sufficiently dispersive bands such that
the low-temperature magnetocaloric properties at the transition
to saturation in some quantum spin chains can be understood
in terms of free spinless fermions \cite{ZhiHo}. Although there
is no such direct connection for flat bands
or in higher dimensions, we may still expect that free spinless fermions
capture relevant qualitative features. This motivates us to analyze
low-energy properties of free spinless fermions on the sawtooth
chain (section \ref{secSaw}) and the kagom\'e lattice (section \ref{secKag}),
after a brief summary of some basic equations in section \ref{secBasic}.
In section \ref{secHub} we discuss implications for the
Hubbard model on the same lattices and mention a relation to
flat-band ferromagnetism \cite{Mielke92,Tasaki92,MieTa,Tasaki98,NGK}.

\section{Model and basic equations}

\label{secBasic}

A central quantity of this paper is the entropy $S$.
It is given in terms of the partition function $Z$
by\footnote{The Boltzmann constant will be set to unity
throughout this paper $k_B = 1$.}
\begin{equation}
S(T) = {\partial \over \partial T} \, T \, \ln Z \, .
\label{defEnt}
\end{equation}
Furthermore, we will analyze free spinless
fermions, whose Hamiltonian reads
\begin{equation}
H = \sum_{\langle i, j\rangle} t_{i,j} \, \left(
\hat{c}^{\dagger}_i \hat{c}_j + \hat{c}^{\dagger}_j \hat{c}_i\right)
+ \mu \sum_{i=1}^N \hat{n}_i \, .
\label{HfreeF}
\end{equation}
The first sum runs over the nearest-neighbor pairs
$\langle i, j\rangle$ of a lattice with $N$
sites. $\hat{c}^{\dagger}_i$ and $\hat{c}_i$ are fermion
creation and annihilation operators at site $i$
with anticommutation relations
$\left\{ \hat{c}^{\dagger}_i,\hat{c}_j\right\} = \delta_{i,j}$,
and $\hat{n}_i = \hat{c}^{\dagger}_i \hat{c}_i$ is the number operator
at site $i$. The hopping parameter $t_{i,j} > 0$
corresponds to the exchange constant of an antiferromagnetic
spin-$s$ $XXZ$ model, and the chemical potential
$\mu$ to a magnetic field $h$ along the $z$-axis
(the sign of $\mu$ has been chosen positive in order to
allow for a direct comparison with the $XXZ$ model in
a magnetic field). We denote
the single-fermion energies of (\ref{HfreeF}) by $\varepsilon_k$
and note that this one-particle problem is equivalent to the one-magnon
problem relative to the ferromagnetically polarized state
of the spin-$s$ $XX$ model on the same lattice.
Furthermore, the Hilbert space of spinless fermions
on an $N$-site lattice is isomorphic to that of the $s=1/2$
$XXZ$ model on the same lattice. Nevertheless,
the many-particle states are not equivalent.
Apart from the different statistics, interactions
are absent in the model of spinless fermions (\ref{HfreeF}).
The latter property makes it straightforward
to derive the thermodynamics, which is a
challenging problem for the $XXZ$ model.

For a free Fermi-gas the grand partition
function\footnote{The chemical potential $\mu$ is included in the
definition of the one-particle energy $\varepsilon_k$ for state $k$.}
is given by
$Z = \prod_k \left(1 + {\rm e}^{- \varepsilon_k/T}\right)$.
Substitution into (\ref{defEnt}) yields
the entropy for a system of free fermions
\begin{equation}
S(T) = \sum_k \ln\left(1 + {\rm e}^{- \varepsilon_k/T} \right)
   + {1 \over T} \sum_k {\varepsilon_k \, {\rm e}^{- \varepsilon_k/T} \over
   1 + {\rm e}^{- \varepsilon_k/T}  } \, .
\label{EntFreeF}
\end{equation}

\section{Sawtooth chain}

\label{secSaw}

\begin{figure}[t]
%\vspace{-2ex}%
\centerline{\includegraphics[width=0.5\columnwidth]{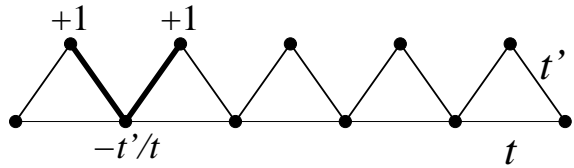}}
\caption{The sawtooth chain. The hopping parameter
along the baseline is $t$, while along the diagonal
directions it is $t'$. For $t' = \sqrt{2} \, t$ the
one-particle states can be localized, as indicated
by the bold line and the coefficients of the wave function
(\ref{locWavSaw}) written next to the three associated sites.}
\label{figSawtooth}
\end{figure}

The first model which we will discuss is the sawtooth chain
shown in Fig.\ \ref{figSawtooth}. The one-particle problem
for (\ref{HfreeF}) on the sawtooth chain leads to a $2 \times 2$
matrix whose eigenvalues yield two bands of one-particle energies
\begin{equation}
\varepsilon_{\pm}(k) = t\, \cos(k) \pm
\sqrt{t^2 \, \cos^2(k) + 2\, t'^2 \, \cos(k) + 2 \, t'^2}
+ \mu \, .
\label{sawFreeFgen}
\end{equation}
When one chooses hopping parameters along the diagonals
$t' = \sqrt{2} \, t$ in terms of the hopping parameters
along the baseline $t$, the one-particle energies
become
\begin{equation}
\varepsilon_{-}(k) = -2 \, t + \mu \, , \qquad
\varepsilon_{+}(k) = 2 \, t \, \left(1 + \cos(k) \right) + \mu \, .
\label{sawFreeFspec}
\end{equation}
Note that the lowest band $\varepsilon_{-}(k)$ is completely flat.
The result (\ref{sawFreeFspec}) is equivalent to the one-magnon
energies of an $XX$ spin model on the sawtooth lattice \cite{jump}.
Very similar results are also obtained for the Heisenberg
model on this lattice (compare eq.\ (33) of \cite{ZhiHo} for
an explicit expression of the one-magnon energies for $s=1/2$).

Now we will concentrate on the case $t' = \sqrt{2}\,t$ where
the lowest band $\varepsilon_{-}(k)$ is completely flat.
As in the $XXZ$ model \cite{jump,RSHSS04}, it is possible to localize
these one-particle states on the three sites indicated
by the bold line (a ,,valley'') in Fig.~\ref{figSawtooth}.
To be more precise,
let us number the sites along the baseline by $2\,i$ and the ones
at the top of the triangles by $2\,i+1$. If we further denote
the fermionic vacuum by $\vert 0 \rangle$, the localized
wave functions are given by (up to normalization)
\begin{equation}
\vert l_{2\,i} \rangle
 = L^\dagger_{2\,i} \, \vert 0 \rangle \, , \qquad
L^\dagger_{2\,i} =
\hat{c}^\dagger_{2\,i-1} + \hat{c}^\dagger_{2\,i+1}
-\sqrt{2} \, \hat{c}^\dagger_{2\,i}  \, .
 \label{locWavSaw}
\end{equation}
At $\mu = 2\,t$ we have $\varepsilon_{-}(k) = 0$. This implies
a degeneracy of the ground states with $0 \le n \le 1/2$
due to the $2^{N/2}$ distinct ways to fill the zero-energy band.
In this manner we have identified a zero-temperature entropy
at $\mu = 2\,t$, $t' = \sqrt{2} \, t$
\begin{equation}
{S(0) \over N} = {\ln{2} \over 2} = 0.34657\ldots \, .
\label{sawT0ent}
\end{equation}
This corresponds to the zero-temperature entropy found
in the Heisenberg model at the saturation field. However, the
value in the latter case is smaller, namely
$S(0)/N = 0.2406059\ldots$ \cite{ZhiHo,DerRi,ZhiTsu}.
That a larger value is obtained for free fermions
can be understood as follows. Many-particle states
can be constructed from localized one-particle states
in different valleys. For spinless fermions it is
possible to put one-particle states in neighbouring
valleys. The corresponding two-particle state
$L^\dagger_{2\,(i+1)} \, L^\dagger_{2\,i} \, \vert 0 \rangle$
is an exact eigenstate with energy $2 \, \varepsilon_{-}$.
However, in the $XXZ$ model the energy of two localized
magnons occupying neighbouring valleys is higher
since they share a site \cite{ZhiHo,ZhiTsu}. This
condition that two localized magnons may not
occupy two neighboring valleys lowers the ground state
degeneracy and thus the zero-temperature entropy
of the $XXZ$ model.

\begin{figure}[t]
%\vspace{-2ex}%
\centerline{\includegraphics[width=0.78\columnwidth]{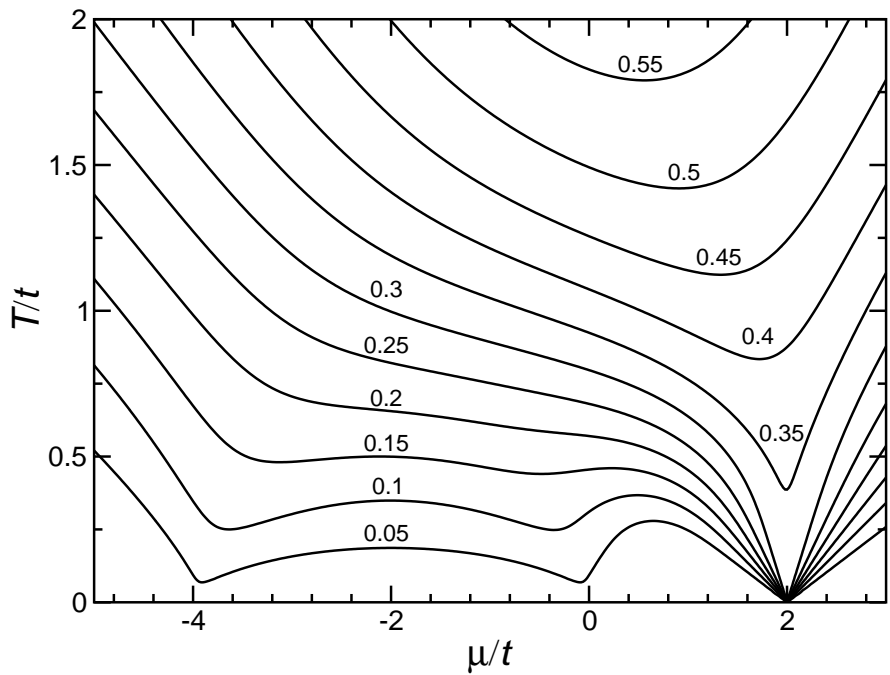}}
\caption{Lines of constant entropy for free spinless fermions
on the sawtooth chain
with $t' = \sqrt{2} \, t$. The value of the entropy per
site $S(T)/N$ is indicated next to each line.}
\label{figSawEnt}
\end{figure}

The entropy at finite temperature and with arbitrary $\mu$
can be evaluated from (\ref{EntFreeF}). One can either
go to the thermodynamic limit and replace the sums by
integrals. Alternatively, one can choose a large system
size ($N=2048$, for example) and evaluate the sums numerically.
The latter has been performed to obtain Fig.~\ref{figSawEnt}
which is indistinguishable from the thermodynamic limit
$N \to \infty$. The right half of this figure can be
compared to Fig.~6(a) and Fig.~7 of \cite{ZhiHo}
which show numerical results for the entropy of the $s=1/2$
Heisenberg model in the $h$-$T$-plane. The following
different low-temperature regions can be distinguished
in Fig.~\ref{figSawEnt}: (i) for $\mu > 2\,t$ both
bands are empty and all excitations
are gapped, (ii) for $\mu = 2\,t$ the ground state is
degenerate with the $T=0$ entropy given in (\ref{sawT0ent}),
(iii) for $2\,t > \mu > 0$ the chemical potential is in
the band gap between the lower band $\varepsilon_{-}(k)$
and the upper band $\varepsilon_{+}(k)$ such that excitations
again have a gap,
(iv) for $0 \ge \mu \ge -4\,t$ the chemical potential is in
the upper band $\varepsilon_{+}(k)$ and one has gapless
low-energy excitations,
(v) for $\mu < -4\,t$ both bands are filled and all excitations
are again gapped.

Now let us start at a small temperature $T$ and change
the chemical potential adiabatically. Such an adiabatic
process keeps the entropy constant, therefore the
temperature $T$ has to change along curves as in Fig.~\ref{figSawEnt}.
If one approaches region (iv) either from region (iii) or
region (v) one sees that temperature drops as
$\mu \to 0$ or $\mu \to -4\,t$. An even larger temperature
drop is observed when one approaches point (ii) either
from region (i) or region (iii). This enhanced cooling
is clearly due to the zero-temperature entropy (\ref{sawT0ent})
at $\mu = 2\,t$. In fact, if the entropy at the starting
point is below this value, one theoretically approaches
$T \to 0$ as $\mu \to 2\,t$. The qualitative behavior
is similar to that of the
$XXZ$ model on the sawtooth chain \cite{ZhiHo} (an interacting
model) despite quantitative differences.

\section{Kagom\'e lattice}

\label{secKag}

\begin{figure}[t]
%\vspace{-2ex}%
\centerline{\includegraphics[width=0.52\columnwidth]{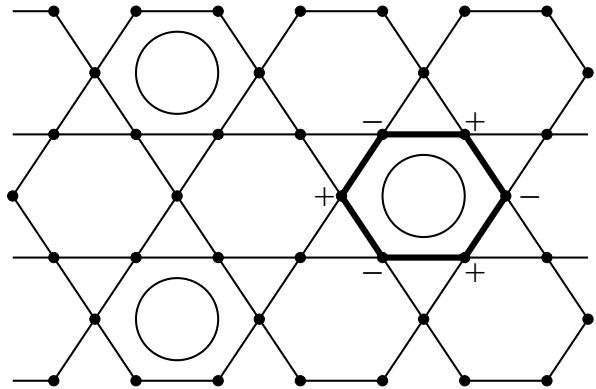}}
\caption{The kagom\'e lattice. The bold hexagon indicates a localized
one-particle state, the signs in the corresponding wave
function are written next to the six sites of this hexagon.
Circles indicate the closest non-overlapping packing of
localized one-particle states.}
\label{figKagome}
\end{figure}

\begin{figure}[t]
%\vspace{-2ex}%
\centerline{\includegraphics[width=0.7\columnwidth]{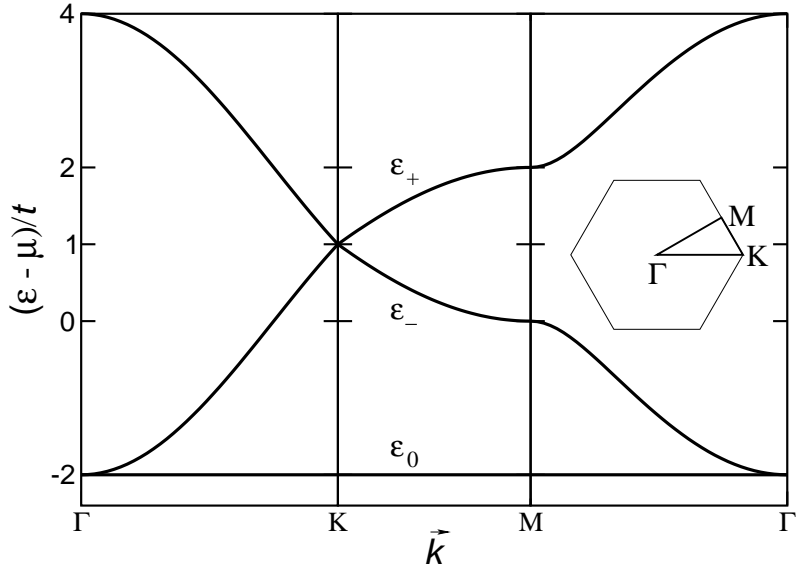}}
\caption{The three bands $\varepsilon_i(\vec{k})$ of single-particle
energies for spinless fermions on the kagom\'e lattice along the path
in the Brillouin zone shown in the inset.
Note that $\varepsilon_0(\vec{k})$ is completely independent of $\vec{k}$.}
\label{figKenergy}
\end{figure}

In this section
we discuss a two-dimensional lattice, the kagom\'e lattice shown in
Fig.~\ref{figKagome}. Since a unit cell of this lattice contains three sites,
the one-particle problem for the Hamiltonian (\ref{HfreeF})
now leads to a $3 \times 3$ matrix whose
eigenvalues yield three bands of one-particle energies
\begin{eqnarray}
\varepsilon_{0}(k_x,k_y) &=& \mu -2 \, t \, , \nonumber \\
\varepsilon_{\pm}(k_x,k_y) &=& \mu + t \pm t \, \sqrt{
1 + 4 \, \cos\left({k_x \over 2}\right) \, \left(
\cos\left({\sqrt{3} \, k_y \over 2}\right) + \cos\left({k_x \over 2}\right)
 \right)} \, .
\label{kagfreeFspec}
\end{eqnarray}
Fig.~\ref{figKenergy} shows these three one-particle bands along
a path in the Brillouin zone. Note that the lowest band $\varepsilon_0(\vec{k})$
turns out to be completely flat.
These results can be related to known ones. Firstly, using a
particle-hole transformation \cite{Mielke92}, the present
problem maps to a tight-binding problem of electrons moving on
the kagom\'e lattice (compare Fig.~\ref{figKenergy} e.g.\ with
Fig.~1(b) of \cite{kTight}). The one-magnon problem for the
$XXZ$ model on the kagom\'e lattice is another equivalent problem
(see \cite{jump} and in particular Fig.~2.31 of \cite{RSH04},
compare also eq.\ (1) of \cite{ZhiTsu}).

As in the $XXZ$ model \cite{jump,RSH04}, it is again
possible to localize these one-particle states in real space,
namely on a hexagon, as indicated by the bold line in Fig.~\ref{figKagome}.
If we number the sites around a hexagon $h$ with $j=1$, \ldots, $6$,
these localized wave functions can be written as (compare also
Fig.~\ref{figKagome} for the signs $(-1)^j$)
\begin{equation}
\vert l_h \rangle
 = L^\dagger_h \, \vert 0 \rangle \, , \qquad
L^\dagger_h = {1 \over \sqrt{6}}
 \sum_{j=1}^6 (-1)^j \, \hat{c}^\dagger_{h,j} \, .
\label{locWavKag}
\end{equation}
The energy of the lowest band becomes $\varepsilon_{0}(\vec{k}) = 0$
at $\mu = 2\,t$. Consideration of all possible ways to fill this
zero-energy band yields $2^{N/3}$ zero-energy excitations. For $\mu = 2\,t$
another one-particle energy vanishes, namely $\varepsilon_{-}(0,0) = 0$
(see Fig.~\ref{figKenergy}). However,  this occurs only at
a single point $\vec{k} = \vec{0}$. Therefore, the additional contribution
can be neglected in the thermodynamic limit $N \to \infty$. Hence, the
zero-temperature entropy per site is at $\mu = 2\,t$
\begin{equation}
{S(0) \over N} = {\ln{2} \over 3} = 0.231049\ldots \, .
\label{kagT0ent}
\end{equation}
The $XXZ$ model on the kagom\'e lattice also gives rise to a
finite zero-temperature entropy at the saturation field
\cite{RSH04,DerRi,ZhiTsu}. A mapping of the configurations
(\ref{locWavKag}) to the exactly solved problem
of hard hexagons \cite{BaTsa,Baxter}
yields a lower bound $S(0)/N \ge 0.11108$ in this case \cite{DerRi,ZhiTsu}.
On the one hand, one expects to find a lower entropy in the $XXZ$
model than for spinless fermions, because creation of localized
magnons on two neighbouring hexagons costs no energy for free spinless
fermions while localized magnons can be created for the $XXZ$ model
only on spatially separated hexagons. On the other hand, the situation remains
more complicated than in one dimension because of additional topological
effects. In fact, a class of many-magnon states has been found
in the $XXZ$ model which are independent of hard-hexagon configurations
\cite{ZhiTsu,ZhiTsuP}. Therefore we believe that the precise value
of the zero-temperature entropy at the saturation field of the $s=1/2$
Heisenberg model on the kagom\'e lattice deserves further attention.

\begin{figure}[t]
%\vspace{-2ex}%
\centerline{\includegraphics[width=0.78\columnwidth]{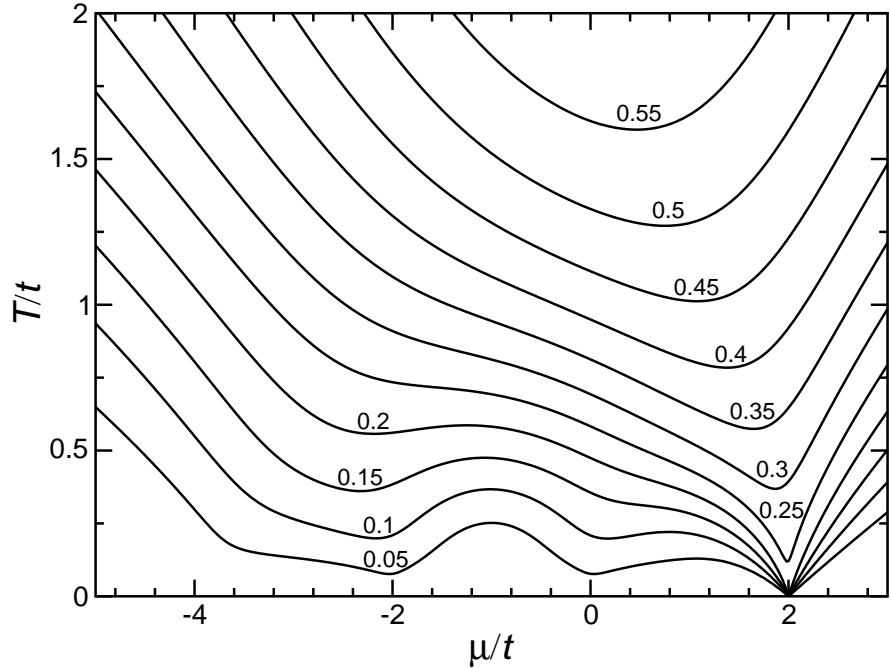}}
\caption{Lines of constant entropy for free spinless fermions on
the kagom\'e lattice. The value of the entropy per
site $S(T)/N$ is indicated next to each line.}
\label{figKagEnt}
\end{figure}

Also for the kagom\'e lattice it is straightforward to obtain
the entropy at finite temperature and with arbitrary $\mu$
from (\ref{EntFreeF}) and the single-particle energies (\ref{kagfreeFspec}).
We have again evaluated the sums numerically on
a large but finite lattice ($N=10800$). The corresponding values
of the entropy per site are shown in Fig.\ \ref{figKagEnt} and
can be considered as representative for the thermodynamic limit.
The following four low-temperature regions can be distinguished
in  Fig.~\ref{figKagEnt}: (i) for $\mu > 2\,t$ all
bands are empty and all excitations
are gapped, (ii) for $\mu = 2\,t$ the ground state is
degenerate with the $T=0$ entropy given in (\ref{kagT0ent}),
(iii) for $2\,t \ge \mu \ge -4\,t$ the chemical potential is in
one of the bands $\varepsilon_{\pm}(\vec{k})$ and one has gapless
low-energy excitations,
(iv) for $\mu < -4\,t$ all bands are filled and all excitations
are again gapped. Additional structures in region (iii) can be
attributed to features in the $T=0$ single-particle density
of states (compare 
Fig.~\ref{figKenergy}). Note that in contrast to the sawtooth
chain, the kagom\'e lattice does not give rise
to a gap for $\mu$ slightly below $2\,t$.

If one now changes $\mu$ adiabatically, temperature drops
considerably as $\mu \to 2 \, t$. One can even reach $T=0$
during an adiabatic process
if the starting entropy is less than the $T=0$ entropy
(\ref{kagT0ent}) at $\mu = 2\, t$. Note the asymmetry of the curves
in Fig.~\ref{figKagEnt} around $\mu = 2\, t$: The dependence on
$\mu$ is generally weaker in the gapless regime (iii) than in the
gapped regime (i). One also observes a cooling effect as
$\mu \to -4\,t$ from the gapped regime (iv), although unlike in
Fig.~\ref{figSawEnt} there is no pronounced minimum close to
$\mu = -4\, t$.

\section{Hubbard model and flat-band ferromagnetism}

\label{secHub}

Now we will discuss the influence of interactions using
the Hubbard model as example. The fermions are assigned
an additional spin index $\sigma = \uparrow$, $\downarrow$,
and there is an on-site Coulomb repulsion $U > 0$ for fermions
(electrons) with different spin:
\begin{equation}
H = \sum_{\sigma} \sum_{\langle i, j\rangle} t_{i,j} \, \left(
\hat{c}^{\dagger}_{i,\sigma} \hat{c}_{j,\sigma}
 + \hat{c}^{\dagger}_{j,\sigma} \hat{c}_{i,\sigma}\right)
+ U \sum_{i=1}^N \hat{n}_{i,\uparrow} \hat{n}_{i,\downarrow}
+ \mu \sum_{\sigma} \sum_{i=1}^N \hat{n}_{i,\sigma} \, .
\label{Hhubbard}
\end{equation}
The Coulomb repulsion is not effective
for spin-polarized configurations (all $\sigma$ equal). These
configuration can be identified with spinless fermions. Conversely,
each eigenstate of the spinless fermion Hamiltonian (\ref{HfreeF})
yields a spin-polarized eigenstate of the Hubbard model
(\ref{Hhubbard}). In particular the one-particle problem is
identical in both models. The single-particle bands
of the Hubbard model are simply twofold degenerate.
Using results from the
previous sections, we can now establish a $T=0$ entropy for
the Hubbard model (\ref{Hhubbard}) for $U>0$ and otherwise the
same conditions which lead to $T=0$ entropy for spinless fermions
(\ref{HfreeF}). A relation to flat-band ferromagnetism
\cite{Mielke92,Tasaki92,MieTa,Tasaki98,NGK} is also evident.

Let us assume that the lowest single-particle band is completely flat,
as we found for the sawtooth lattice (Fig.~\ref{figSawtooth}) at
$t' = \sqrt{2} \, t$ and the kagom\'e lattice (Fig.~\ref{figKagome}).
The corresponding states can be localized in real space as before;
the creation operator of a local state $L^\dagger_{j,\sigma}$ just
acquires an additional spin index. States
localized at $j_1$, $j_2$, \ldots with
spins $\sigma_1$, $\sigma_2$, \ldots  can be created by repeated
application of distinct creation operators
\begin{equation}
L^\dagger_{j_1,\sigma_1} L^\dagger_{j_2,\sigma_2} \cdots 
 \vert 0 \rangle \, .
\label{Lmany}
\end{equation}
These states are clearly eigenstates for any $j_1 \ne j_2$ if they
are spin-polarized $\sigma_j = \sigma$.
As long as the local states are created in spatially
non-overlapping regions, the states (\ref{Lmany}) with
$\sigma_1 \ne \sigma_2$ are also eigenstates with the same energy as the
spin-polarized states. However, once the localization
regions overlap, there are generally two electrons with distinct
spin at a site. The Coulomb repulsion
in (\ref{Hhubbard}) raises the energy of such states by an amount
$\propto U$. Note that certain combinations where local
states with different spins overlap must have the same energy
as the spin-polarized states because of the $SU(2)$-symmetry of the
Hamiltonian (\ref{Hhubbard}). However, if Coulomb repulsion is active,
it will raise the energy of many-particle states in the
lowest band.

Now choose $\mu$ such that the lowest band has zero energy
$\varepsilon(\vec{k})=0$ ($\mu = 2\,t$ for the sawtooth chain and the
kagom\'e lattice). The preceding discussion then yields a lower
bound of the $T=0$ entropy of the Hubbard model (\ref{Hhubbard})
in terms of spinless fermions (\ref{HfreeF}): $S_{\rm Hubbard}(0) \ge
S_{SF}(0)$. The precise value of the entropy of the Hubbard model
(\ref{Hhubbard}) is difficult to determine because of the additional
zero-energy states which are not spin-polarized. In any case,
also in the Hubbard model adiabatic changes of $\mu$ will give
rise to large temperature variations.

Spatial overlaps of the individual local states cannot be avoided
at large filling fractions of the flat band. Spin-polarized states
still have the lowest possible energy in this case while a non-polarized
state is generally pushed higher in energy by Coulomb repulsion.
This implies a phenomenon known as flat-band ferromagnetism
\cite{Mielke92,Tasaki92,MieTa,Tasaki98,NGK}:
In this region of densities spin-polarized, i.e.\ ferromagnetic
configurations are preferred over non-magnetic
configurations. To be more precise, let us consider the
kagom\'e lattice. Using our conventions (\ref{Hhubbard}),
the lowest band can be filled in a spin-polarized manner for
$0 \le n \le 1/3$. On the other hand, local wave functions have
to overlap for $n > 1/9$ (see Fig.\ \ref{figKagome}). Slightly above
the threshold $n=1/9$, ferromagnetism can still be avoided by having overlap
only for localized states with equal spins. According to a detailed
mathematical analysis \cite{Mielke92}, the ground state of the
Hubbard model (\ref{Hhubbard}) is ferromagnetic for $1/6 < n \le 1/3$.

\section{Summary and Outlook}

In sections \ref{secSaw} and \ref{secKag}
we have examined spinless fermions on the sawtooth
chain and the kagom\'e lattice. We found that the lowest single-particle
band is completely flat for a particular choice of hopping
parameters in the case of the sawtooth chain and without any
fine-tuning for the kagom\'e lattice. These states can be
localized in real space. Since the particles have zero
energy at $\mu = 2\, t$, one finds a finite $T=0$ entropy at this value
of the chemical potential. This implies large adiabatic temperature
variations when $\mu$ is changed adiabatically at finite temperature.
At least on the single-particle level, the construction
is completely analogous
to the $XXZ$ model on the same lattice. It can therefore be
generalized to many other lattices for which localized magnons
have been constructed \cite{jump,RSHSS04,RSH04,RiLoc},
including in particular the pyrochlore lattice in three dimensions.
We do not wish to speculate about possible experimental applications.
Rather we regard these non-interacting models as an illustration of
important qualitative features of spin models. First experimental
investigations \cite{ZGBP,SPSGBPBZ} of the magnetocaloric effect in
highly frustrated magnets shows that they are indeed promising
candidates for efficient low-temperature magnetic cooling.

Tuning of the chemical potential $\mu$ through a
non-degenerate band-edge, i.e.\ an edge without flat directions,
yields a second-order quantum phase transition. Spinless fermions
thus provide a simple realization of the quantum critical scenario
discussed in detail in \cite{ZGRS,GaRo}.

In section \ref{secHub} we have briefly commented on the Hubbard
model where an on-site Coulomb repulsion $U$ gives rise to
interactions. However, this repulsion is not active for
spin-polarized configurations. Therefore,
the existence of a finite $T=0$ entropy for a certain value of $\mu$
and consequently the large temperature variations when $\mu$ is
tuned through this point carry over from spinless fermions to the
Hubbard model. Nevertheless, both the $T=0$ entropy and
the cooling rate in the Hubbard model remain to be analyzed
in more detail. In this context it is also
intriguing to note that a finite $T=0$ entropy arises in the
Hubbard model under the same conditions which lead to
flat-band ferromagnetism \cite{Mielke92,Tasaki92,MieTa,Tasaki98,NGK}.

\section*{Acknowledgements}

The present work was inspired by discussions  and collaborations with
H.\ Frahm and M.E.\ Zhitomirsky which are gratefully acknowledged.


\begin{thebibliography}{99}
\bibitem{lhuillier03} Misguich~G., Lhuillier~C.,
              Two-dimensional quantum antiferromagnets,
% preprint cond-mat/0310405.
              p.~229--306
              in {\it Frustrated Spin Systems}. Ed.\
              Diep~H.T., World Scientific, Singapore, 2004.
\bibitem{RSH04} Richter~J., Schulenburg~J., Honecker~A.,
              Lect.\ Notes Phys., 2004, {\bf 645}, 85.
\bibitem{jump} Schulenburg~J.\ et al.,
% Honecker~A., Schnack~J., Richter~J., Schmidt~H.-J., 
	      Phys.\ Rev.\ Lett., 2002, {\bf 88}, 167207.
\bibitem{RSHSS04} Richter~J.\ et al.,
% Schulenburg~J., Honecker~A., Schnack~J., Schmidt~H.-J.,
              J.\ Phys.: Condens.\ Matter, 2004, {\bf 16}, S779.
\bibitem{SSRS01} Schnack~J., Schmidt~H.-J., Richter~J., Schulenburg~J., Eur.\
              Phys.\ J. B, 2001, {\bf 24}, 475.
\bibitem{RiLoc} Richter~J., % Preprint arXiv:cond-mat/0503243, 2005.
              Low Temperature Physics, 2005, {\bf 31}, 695
\bibitem{Zhito} Zhitomirsky~M.E., Phys.\ Rev.\ B, 2003, {\bf 67}, 104421.
\bibitem{ZGBP} Zhitomirsky~M.E., Golov~A.I., Berkutov~I.B., Petrenko~O.A.,
              Preprint, 2001 \\ \phantom{X} \qquad
              [{\tt http://grendel.ph.man.ac.uk/Research/GGG1.pdf}].
\bibitem{SPSGBPBZ} Sosin~S.S.\ et al.,
% Prozorova~L.A., Smirnov~A.I., Golov~A.I.,  Berkutov~I.B.,
% Petrenko~O.A., Balakrishnan~G., Zhitomirsky~M.E.,
              Phys.\ Rev.\ B, 2005, {\bf 71}, 094413.
\bibitem{ZhiHo} Zhitomirsky~M.E., Honecker~A., J.\ Stat.\ Mech.: Theor.\ Exp.,
              2004, P07012.
\bibitem{DerRi} Derzhko~O., Richter~J., Phys.\ Rev.\ B, 2004, {\bf 70},
              104415.
\bibitem{ZhiTsu} Zhitomirsky~M.E., Tsunetsugu~H., Phys.\ Rev.\ B, 2004,
              {\bf 70}, 100403(R).
\bibitem{Schnack} Schnack~J., Preprint arXiv:cond-mat/0501625, 2005.
\bibitem{ZhiTsuP} Zhitomirsky~M.E., Tsunetsugu~H.,
              Preprint arXiv:cond-mat/0506327, 2005.
\bibitem{HoWe} Honecker~A., Wessel~S., Preprint arXiv: cond-mat/0508123, 2005,
              to appear in Physica~B.
%\\ \phantom{X} \qquad
%              [{\tt http://www.tu-bs.de/\~{ }honecker/papers/sces05.ps.gz}].
\bibitem{takahashi} Takahashi~M., Thermodynamics of one-dimensional solvable
              models. Cambridge University Press, Cambridge, 1999.
\bibitem{Mielke92} Mielke~A., J.\ Phys.\ A: Math.\ Gen., 1992, {\bf 25},
              4335.
\bibitem{Tasaki92} Tasaki~H., Phys.\ Rev.\ Lett., 1992, {\bf 69}, 1608.
\bibitem{MieTa} Mielke~A., Tasaki~H., Commun.\ Math.\ Phys., 1993, {\bf 158},
              341.
\bibitem{Tasaki98} Tasaki~H., Progr.\ Theor.\ Phys., 1998, {\bf 99}, 489.
\bibitem{NGK} Nishino~S., Goda~M., Kusakabe~K., J.\ Phys.\ Soc.\ Jpn., 2003,
              {\bf 72}, 2015.
\bibitem{kTight} Kimura~T., Tamura~H., Shiraishi~K., Takayanagi~H., Phys.\
              Rev.\ B, 2002, {\bf 65}, 081307(R).
\bibitem{BaTsa} Baxter~R.J., Tsang~S.K., J.\ Phys.\ A: Math.\ Gen., 1980,
              {\bf 13}, 1023.
\bibitem{Baxter} Baxter~R.J., J.\ Phys.\ A: Math.\ Gen., 1980, {\bf 13}, L61.
\bibitem{ZGRS} Zhu~L., Garst~M., Rosch~A., Si~Q., Phys.\ Rev.\ Lett., 2003,
              {\bf 91}, 066404.
\bibitem{GaRo} Garst~M., Rosch~A., Preprint arXiv:cond-mat/0506336, 2005.



%\end{thebibliography}

%
%% If you have problems with typesetting in ukrainian uncomment lines below.
%

\label{last@page}

\end{thebibliography}
\end{document}